\documentstyle[epsf,12pt]{article}

\renewcommand{\theequation}{\arabic{equation}}

\newcommand{\beq}{\begin{eqnarray}}
\newcommand{\eeq}{\end{eqnarray}}

\newcommand{\lang}{\left\langle}
\newcommand{\rang}{\right\rangle}
\def\coup{\frac{\alpha_s}{\pi}}
\def\qvec{\mbox{\boldmath $q$}}
\def\pvec{\mbox{\boldmath $p$}}

\def\0vec{\mbox{\boldmath $0$}}
\makeatletter
\def\slash#1{{\mathpalette\c@ncel{#1}}} 
\makeatother

\begin{document}

\begin{titlepage}
%\begin{flushright}
%\begin{tabular}{l}
%\end{tabular}
%\end{flushright}
\vskip0.5cm

\begin{center}
{\Large \bf 
           $J/\psi$-Nucleon Scattering Length and In-medium \\
             Mass Shift of $J/\psi$ in QCD Sum Rule Analysis
  \\}

\vspace{1cm}

\renewcommand{\thefootnote}{\fnsymbol{footnote}}
 {\sc Arata~Hayashigaki}
\footnote[2]{E-mail: arata@yukawa.kyoto-u.ac.jp}
\\[0.3cm]

\vspace*{0.1cm} {\it Yukawa Institute for Theoretical Physics,
Kyoto University, Kyoto 606-8502 Japan}
\\[1cm]

  \vskip1.8cm
  {\large\bf Abstract:\\[10pt]} \parbox[t]{\textwidth}{

We calculate the spin-averaged $J/\psi$-nucleon scattering length 
$a_{J/\psi}$ by directly applying QCD sum rule to $J/\psi$-$N$ forward
scattering amplitude. Our result $a_{J/\psi} = - 0.10 \pm 0.02$ fm,
predicts the possibility of bound states with nuclei,
though the force is weaker than that  
of the light vector mesons ($\rho,\omega,\phi$)-$N$ cases.
Up to dimension-4 gluonic operators, we evaluate the scattering length 
with twist-2 contribution, which increases the absolute value of 
the scattering length about $30\%$.
If we apply $a_{J/\psi}$ to the effective mass 
of $J/\psi$ in nuclear 
matter on the basis of the linear density approximation, 
it shows very slight decrease 
($4 \sim 7$ MeV) at normal matter density.
}
\end{center}

\vskip1cm


\noindent
[Keywords: QCD sum rule, moment sum rule, OPE, finite nucleon density, 
scattering length, effective mass, heavy quark, $J/\psi$, twist-2 operator]

\vskip1cm 

\end{titlepage}

\section{Introduction}
\setcounter{equation}{0}
\renewcommand{\theequation}{\arabic{section}.\arabic{equation}}
Theoretical analysis on the in-medium properties of hadrons
is increasingly 
required by various on-going and forthcoming heavy-ion experiments
(such as SPS, LHC (CERN) and AGS, RHIC (BNL)) \cite{QM97}. 
In particular, experimentally it is important 
to observe vector mesons, because they
decay into lepton pairs and carry the information inside the matter
without disturbance of the strong interaction. 
The properties of light vector mesons in nuclear matter have been
extensively studied in various theoretical approaches such as
effective hadronic models \cite{BR} 
and QCD sum rules (QSR's) \cite{HL,YK,JL,KH}.
Vacuum properties of the vector mesons have been successfully
studied by the QSR's \cite{SVZ,RRY1}. The method enables us to express
physical quantities such as mass and decay width in terms of
the parameters in QCD Lagrangian and vacuum condensates. 
Extending the vacuum QSR to finite density, we can consistently
incorporate the effects of nuclear matter 
into the form of in-medium condensates.
There are two methodologically different ways for in-medium QSR.
Firstly, T. Hatsuda and S.H. Lee developed the in-medium QSR formalism 
for light vector mesons \cite{HL}. They found a $10 \sim 20\;\%$ decrease 
of the $\rho$ and $\omega$ mesons at normal matter density. 
Secondly, for light vector mesons 
we formulated in-medium QSR \cite{YK,KH}
based on the relation between a scattering length 
and a mass shift \cite{KM}. In this approach with the Fermi gas model,
in-medium correlation function is devided into vacuum part and
one nucleon part. This one nucleon part corresponds to 
the forward vector meson-nucleon scattering amplitude.
The QSR analysis on the forward scattering amplitude
enables us to obtain the information for vector meson-nucleon interaction.
Moreover, from the information we can estimate the change of spectra
for vector mesons in nuclear matter.
The difference between these two approaches 
has been discussed in \cite{HLS,KH}. 
Eventually we derived in \cite{KH} that 
both of them are based on almost the same idea and can 
lead the consistent results
with those of the effective models. 

In this paper we apply the QSR analysis established in \cite{KH} 
to a heavy quark system with equal mass for quark and antiquark.
As a concrete system we focus on $J/\psi$, which is a
low-lying charmonium state ($^3S_1$). 
To study the medium modification of $J/\psi$ has the following reasons:
\begin{enumerate}
\item 
We have the detailed experimental information 
for the charmonium. In particular, the spectrum of $J/\psi$ is 
extremely narrow for the leptonic decay 
($\Gamma_{l^+l^-}$ $\simeq$ $5$ keV).
So it would be good tool to observe
the change of the spectra (e.g., mass shift) in nuclear matter.
\item
Since charmonium and nucleons consist of quarks in different
kinds of flavors, $J/\psi$-$N$ interaction 
is purely gluonic without quark exchange to first order
in elastic scattering. This simplification reduces our
practical calculation.
\item
Theoretical studies for $J/\psi$ in QSR have succeeded only in the
description of the free state \cite{RRY1,RRY2}.
\item 
To utilize $J/\psi$ suppression \cite{MS} as a direct signal 
of the quark-gluon plasma (QGP) phase, we need to estimate
the effect of nuclear absorption theoretically \cite{Khar}.
It seems prompt to conclude that the present experimental data \cite{NAC} 
can be explained only by nuclear absorption, until
we investigate the $J/\psi$-nucleon interaction in detail.
For that purpose, it is reasonable as the first step
to study the $J/\psi$-$N$ elastic scattering at low energy.
\end{enumerate}

Motivated by these,
we calculate $J/\psi$-$N$ scattering length and 
the mass shift of $J/\psi$ in nuclear matter.
That is, the first aim is to estimate 
the essential features of the interaction
between $J/\psi$ and $N$ through the scattering length.
In practice, by applying QSR to $J/\psi$-$N$
forward scattering amplitude we calculate the scattering length.
The scattering length is a physically very important quantity in free space,
because it is the unique observable in $J/\psi$-$N$ elastic scattering
at low energy.
If it is negative, then we could predict attractive nuclear force
capable of binding $J/\psi$ to nucleus, so that 
$J/\psi$ could lead to a bound
state with nucleus. Moreover, due to absence of Pauli
blocking unlike the case of light vector mesons-$N$ system,
the effective $J/\psi$-$N$ interaction will not have a short-range
repulsion. The prediction of such exotic state
will give exciting new directions in nuclear physics.
As is well known,
since the nuclear force is repulsive for isovector mesons,
$\pi$ meson forms $\pi$-nucleus bound state by Coulomb attractive
force.
On the other hand $J/\psi$ is expected to be bounded only by attractive 
nuclear force from the isoscalar property. As was pointed out in \cite{BST}, 
this interaction should be sufficiently attractive to allow 
a bound state. The probability of such exotic states has recently 
discussed for $\eta,\omega$ and $D$ cases \cite{HHG}.

The second aim is how the superposition of elementary $J/\psi$-$N$
scattering at low energy affects the effective mass of $J/\psi$
in nuclear matter.
When we work in the dilute nucleon gas, we find that the mass shift
is linearly dependent on the density (linear density approximation). 

This paper is organized as follows.
In section 2 we summarize the relation between the scattering length and
the mass shift in the linear density
approximation \cite{KH}. In the actual calculation 
we adopt moment sum rule method to the forward scattering amplitude.
In section 3 the Wilson coefficients in the OPE side is explicitly given 
for twist-2 operator.
In section 4 in order to obtain unknown hadronic parameters 
for the forward scattering amplitude
we apply moment sum rule to vacuum correlation function. 
In section 5 the numerical results of the scattering 
length and the mass shift of $J/\psi$ is shown.
Finally concluding remarks are given.

\vskip0.5cm

\section{The relation between 
scattering length and mass shift}
\setcounter{equation}{0}
\renewcommand{\theequation}{\arabic{section}.\arabic{equation}}
Let us first review the relation between the scattering length
and the mass shift on the basis of QSR method \cite{YK,KH}.
The starting point of this approach is 
the following vector current correlation function
in the ground state of 
nuclear matter with nucleon density $\rho_N$. 
\begin{eqnarray}
\Pi_{\mu\nu}^{NM}(q) = i\int d^{4}x e^{iq \cdot x}
\langle  \mbox{T}J_{\mu}(x)
J_{\nu}^{\dag}(0) \rangle _{NM(\rho_N)},
\label{eqn:1}
\end{eqnarray}
where $q^{\mu}=(\omega,\qvec\,)$ is the four-momentum carried by
the $J/\psi$ vector meson current $J_{\mu}(x)$ $=$
$\overline{c}\gamma_{\mu}c(x)$ with the quantum numbers.
Following the QSR method, when we apply an operator product expansion
(OPE) to this correlator at deep Euclidean region
($Q^2=-q^2>0$), it is supposed
that the $\rho_N$-dependence of this correlator is entirely contained
into the $\rho_N$-dependence of various condensates. 
Moreover we assume the Fermi gas model taking into
account the Pauli principle among uncorrelated nucleons 
for the nuclear matter. In this approximation, in-medium 
correlation function reads 
\begin{eqnarray}
\Pi_{\mu\nu}^{NM}(q) &=& \Pi_{\mu\nu}^{0}(q)+
\sum_{spin,isospin}\int^{p_{F}}\frac{d^{3}p}{(2\pi)^{3}2p_{0}}
T_{\mu\nu}(q),
\label{eqn:2}
\end{eqnarray}
where $\Pi_{\mu\nu}^{0}(q)$ is in-vacuum correlation function and
$\sum_{spin,isospin}$ denotes the sum of spin and isospin states 
for nucleons in nuclear matter.	
$T_{\mu\nu}(q)$ is the vector current-nucleon forward scattering
amplitude defined as
\begin{eqnarray}
T_{\mu\nu}(\omega,\qvec\,) =
i\int d^{4}x e^{iq\cdot x}\langle N(ps)|
\mbox{T}J_{\mu}(x)J_{\nu}^{\dag}(0) |N(ps) \rangle.
\label{eqn:3} 
\end{eqnarray}
Here $|N(ps)\rangle$ denotes the nucleon state with four momentum
$p = (p_{0},\pvec)$ and spin $s$ normalized covariantly as 
$\langle N(\pvec)|N(\pvec')\rangle  = (2\pi)^{3}
2p^{0}\delta^{3}(\pvec-\pvec')$. 
$\Pi_{\mu\nu}^{0}(q)$ gives the  main contribution for
$\Pi_{\mu\nu}^{NM}(q)$ due to the perturbative contribution.
On the other hand $T_{\mu\nu}(q)$ leads small contribution for
$\Pi_{\mu\nu}^{NM}(q)$, but the effect is vital contribution. 

If we consider sufficiently low nucleon density such 
as normal matter density 
($\rho_N\sim 0.17$ fm$^{-3}$), the integral 
of the last term in Eq.(\ref{eqn:2}) can be approximated up to
the first order of nucleon density $\rho_N$ reasonably well.
The linear density term corresponds to the matter
with static nucleons ($\pvec={\bf 0}$) and higher order correction terms
correspond to the velocity-dependent terms 
involving the effect of Fermi motion ($\pvec\neq {\bf 0}$) and the complex 
interaction among nucleons. 
The linear expression can be calculated model-independently.
On the other hand the higher order corrections
depend on the model
calculation, but in a few effective theory \cite{CFG}
it is known that the effect for the linear result
is fairly small ($\sim 10\%$) at nuclear matter saturation density.
Hatsuda et al. also insist that
the Fermi momentum correction is fairly small ($\sim 10\%$)  
up to twist-4 operators in \cite{HLS}. 
Thus we can safely neglect the effect at the saturation density. 
Therefore we can set $p=(M_{N},{\bf 0})$ for $T_{\mu\nu}(p,q)$,
so that we proceed to discussions based on the assumption
that all nucleons are at rest in nuclear matter.

In Eq.(\ref{eqn:2}) the second term means the slight
deviation from the properties in free state determined by
$\Pi^0_{\mu\nu}$. By applying QSR method to $T_{\mu\nu}$ directly,
we relate the
scattering length extracted from the QSR for $T_{\mu\nu}$
with the mass shift as one of deviation from the free state
in the framework of QSR.
Near the pole position of the $J/\psi$, 
$T_{\mu\nu}$ can be associated with
the $T$ matrix for the forward $J/\psi$-$N$ helicity amplitude
${\cal T}_{hH,h'H'}(\omega,\qvec\,)$, where $h(h')$ and $H(H')$
are the helicity of the initial(final) $J/\psi$ and
the initial(final) nucleon, respectively.
The relation between $T_{\mu\nu}$ and ${\cal T}_{hH,h'H'}$ is given
by the relation
\begin{eqnarray}
\epsilon^{\mu^{*}}_{(h')}(q)T_{\mu\nu}(\omega, \qvec)
\epsilon^{\nu}_{(h)}(q) \simeq
\frac{-f_{J/\psi}^{2}m_{J/\psi}^{4}}{(q^{2}-m_{J/\psi}^{2}+
i\varepsilon)^{2}}{\cal T}_{hH,h'H'}(\omega,\qvec).
\label{eqn:4}
\end{eqnarray}
Here we introduce the coupling $f_{J/\psi}$
and the $J/\psi$ mass $m_{J/\psi}$ by the relation
$\langle 0|J_{\mu}|J/\psi^{(h)}(q)\rangle$ 
$=$
$f_{J/\psi}m_{J/\psi}^{2}\epsilon_{\mu}^{(h)}(q)$ 
with the polarization vector $\epsilon_{\mu}^{(h)}$
normalized as
$\sum_{h}\epsilon_{\mu}^{(h)^{*}}(q)\epsilon_{\nu}^{(h)}(q)$
$=-$ $g^{\mu\nu}+q^{\mu}q^{\nu}/q^{2}$.
Taking the spin average on both sides of Eq.(\ref{eqn:4}),
$T_{\mu\nu}(\omega,\qvec)$ is projected onto 
$T(\omega,\qvec)=T_{\mu}^{\mu}/(-3)$ and 
${\cal T}_{hH,h'H'}(\omega,\qvec)$ is projected onto 
the spin averaged $J/\psi$-$N$ $T$-matrix, ${\cal T}(\omega,\qvec)$.
At low energy, $q=(m_{J/\psi},{\bf 0})$ and $p=(M_N,{\bf 0})$. 
${\cal T}$ is reduced to the spin averaged $J/\psi$-$N$ scattering length
$a_{J/\psi}=1/3(2a_{3/2}+a_{1/2})$ ($a_{1/2}$ and $a_{3/2}$ are 
the scattering lengths in the spin-1/2 and spin-3/2 channels,
respectively) as
${\cal T}(m_{J/\psi},\qvec={\bf 0}) =
8\pi(M_{N}+m_{J/\psi})a_{J/\psi}$.
We note that the negative $a_{J/\psi}$ corresponds to 
attraction in our convention.

We relate the parameters of the QCD Lagrangian with
the hadronic mass and coupling using the dispersion relation.
If one utilizes the retarded
correlation function as a useful quantity for dispersion analysis,
we obtain the following dispersion relation for $T(\omega,\qvec)$;
\begin{eqnarray}
T(\omega,{\bf 0})=\frac{1}{\pi}\int_{-\infty}^{\infty}du
\frac{\rho(u,{\bf 0})}{u-\omega-i\varepsilon}=
\frac{1}{\pi}\int_{0}^{\infty}du^2
\frac{\rho(u,{\bf 0})}{u^2-\omega^2}.
\label{eqn:5}
\end{eqnarray}
Here the spectral function $\rho(u,\qvec={\bf 0})$ is given 
with three unknown phenomenological parameters $a,b,c$ in terms of
the spin-averaged $J/\psi$-$N$ forward T-matrix ${\cal T}$ 
such as 
\begin{eqnarray}
\rho(u,\qvec={\bf 0}) &=& \frac{1}{\pi}\mbox{Im} \left[
\frac{-f_{J/\psi}^{2}m_{J/\psi}^{4}}{(u^{2}-
m_{J/\psi}^{2}+i\varepsilon)^{2}}{\cal T}(u,{\bf 0}) \right]
+ \cdots 
\label{eqn:6.1} \\
&=& a\,\delta'(u^{2}-m_{J/\psi}^{2}) + b\,\delta(u^{2}-m_{J/\psi}^{2})
 + c\,\delta(u^{2}-s_{0}).
\label{eqn:6.2}
\end{eqnarray}
$\cdots$ term in Eq.(\ref{eqn:6.1}) represents 
the continuum contribution and $\delta'$ in Eq.(\ref{eqn:6.2}) 
is the first derivative of $\delta$ function
with respect to $u^2$. The first $a$-term is the double pole term 
corresponding to the on-shell effect of $T$ matrix and the coefficient
is associated with the scattering length $a_{J/\psi}$ as
$a=8\pi f_{J/\psi}^{2}m_{J/\psi}^{4}(M_{N}+m_{J/\psi})a_{J/\psi}$.
The second $b$-term is the simple pole term corresponding to the off-shell
effect of $T$ matrix. The third $c$-term is the continuum term 
corresponding to other remaining effects, where $s_{0}$ is regarded as
the continuum threshold in vacuum. 
Now the contribution from the inelastic channels is not 
included in the ansatz of Eq.(\ref{eqn:6.2}).
In this system the OZI rule restricts the inelastic channels of
$J/\psi$-$N$ interactions to those containing charmed quarks,
for example, $J/\psi+N$ $\rightarrow$ $D+\bar{D}+N$ and
$J/\psi+N$ $\rightarrow$ $\Lambda_c+\bar{D}$.
But all these processes are forbidden at the threshold.
So fortunately this system is immune from such inelastic contributions. 

The parameters $a,b$ and $c$ in Eq.(\ref{eqn:6.2}) are not completely
independent. That is, among
these parameters we introduce a constraint relation,
which is imposed by low energy theorem for the
$J/\psi$ current-nucleon forward scattering amplitude. 
In the low energy limit $\omega\rightarrow 0$, $T(\omega,{\bf 0})$ become
equivalent to Born term $T^{\rm Born}(\omega,{\bf 0})$, 
which is zero in $J/\psi$-$N$ system. 
Now we get the following constraint relation from the low energy
theorem,
\begin{eqnarray}
\frac{a}{m_{J/\psi}^4}+\frac{b}{m_{J/\psi}^2}+
\frac{c}{s_0}=0.
\label{eqn:7}
\end{eqnarray}
Therefore the spectral function is parametrized with two unknown 
phenomenological parameters $a$ and $b$ by removing $c$ 
from Eq.(\ref{eqn:7}). 
The phenomenological (PH) side for 
$\Pi^{NM}_{\mu\nu}$ can be expressed as the combination of
pole position for $\Pi^{0}_{\mu\nu}$ and $T_{\mu\nu}$ such as 
\begin{eqnarray}
\Pi^{NM}_{\mu\nu} &=& 
\left(\frac{q_{\mu}q_{\nu}}{q^{2}}-g_{\mu\nu}\right)
\left[\frac{F}{m_{J/\psi}^{2}-q^{2}}+
\frac{\rho_N}{2M_{N}}\left\{\frac{a}{(m_{J/\psi}^{2}-q^{2})^{2}} +
\frac{b}{m_{J/\psi}^{2}-q^{2}}\right\}+\cdots\right] \nonumber \\
&\propto& \frac{F+\Delta F}{(m_{J/\psi}^{2}+\Delta m_{J/\psi}^{2}-q^{2})}
 + \cdots,
\label{eqn:8}
\end{eqnarray}
where the pole residue $F$ in $\Pi^0_{\mu\nu}$ is equivalent to
$f_{J/\psi}^2 m_{J/\psi}^4$ and the deviation $\Delta F$
is $\rho_N b/2M_N$.
The quantity expressed as the shift of the squared $J/\psi$ mass
in nuclear matter,
\begin{eqnarray}
\Delta m_{J/\psi}^{2} = 2  m_{J/\psi}\delta m_{J/\psi} = 
\frac{\rho_N}{2M_{N}}\frac{a}{f_{J/\psi}^{2}m_{J/\psi}^{4}}
 = \frac{\rho_N}{2M_{N}}\,\,8\pi(M_{N}+m_{J/\psi})a_{J/\psi}
\label{eqn:9}
\end{eqnarray} 
is proportional to the scattering length $a_{J/\psi}$ through
the double pole term in $T_{\mu\nu}$.
Thus we can calculate the mass shift $\delta m_{J/\psi}$ 
in Eq.(\ref{eqn:9}) from
$a_{J/\psi}$ obtained by QSR for $T_{\mu\nu}$. 

We explicitly write down PH side with the 
unknown parameters $a$ and $b$ for $T(q^2)$ using 
Eq.(\ref{eqn:5}),(\ref{eqn:7}) and (\ref{eqn:8}). 
We take the $n$-th derivative 
with respect to $q^2$ after dividing $T^{\rm ph}$ by $q^2$ 
as follows and define it as $\widehat{T}^{(n)}$.
\begin{eqnarray}
\lefteqn{\frac{1}{n!}\left(\frac{d}{dq^2}\right)^n
\frac{T^{{\rm ph}}(q^2)}{q^2}\ 
\equiv \ \widehat{T}^{(n)\;{\rm ph}}(q^2\;;\;a,b)}
\nonumber\\
&=& \frac{a}{m_{J/\psi}^4}
\left[\ \frac{(n+1)m_{J/\psi}^2}{(m_{J/\psi}^2-q^2)^{n+1}}
+\frac{1}{(m_{J/\psi}^2-q^2)^{n+1}}-\frac{1}{(s_0-q^2)^{n+1}}\ \right]
\nonumber\\
&&+\frac{b}{m_{J/\psi}^2}
\left[\
\frac{1}{(m_{J/\psi}^2-q^2)^{n+1}}-\frac{1}{(s_0-q^2)^{n+1}}\ \right].
\label{eqn:10}
\end{eqnarray}

In order to construct the QSR 
we calculate the $n$-th derivative of OPE side similarly in the next section.

\vskip0.5cm

\section{The calculation of the Wilson coefficients for $T_{\mu\nu}$}
\setcounter{equation}{0}
\renewcommand{\theequation}{\arabic{section}.\arabic{equation}}
Now we give the OPE expression for $T_{\mu\nu}$.
The main task in
OPE side is to calculate the Wilson coefficients 
based on perturbative QCD.
In the case of $J/\psi$ the charmed quark mass is so heavy that
the calculation of the Wilson coefficients
must be carried out explicitly with
the effect of heavy quark mass.
We now expand local operators up to dimension-4 in the OPE side.
Then pure gluonic contributions must be only taken into account 
for the local operators.
Up to this order of the OPE
the nucleon matrix elements of two-gluon operators 
($GG$) are most dominant.
We note that contrary to vacuum QCD sum rule
a new feature in $T_{\mu\nu}$ is 
that the nucleon matrix elements 
survive not only Lorentz
scalar operators but also nonscalar operators.
That is, we must consider new contributions from 
twist-2 operators with two spins for the Wilson coefficients.
In order to calculate the coefficient function,
we adopt the valid and well-known method for massive quarks
propagating through the couple to soft gluons 
working as the external field, namely fixed-point gage method \cite{VAF}.
This gauge condition is expressed as $x^\mu A_\mu^a(x)=0$.
The nucleon matrix element of two-gluon operators can be
decomposed into the scalar part
and the twist-2 part with an additional four-vector $u_\mu$ ($u^2=1$),
through the simple tensor analysis \cite{JCFG}.
\begin{eqnarray}
\lang G_{\alpha\beta}^a G_{\gamma\delta}^b\rang_{N}
&=& \frac{\delta^{ab}}{96}\ \biggl.\biggr[\ \lang G^2\rang_N 
(g_{\alpha\gamma}g_{\beta\delta}-g_{\alpha\delta}g_{\beta\gamma})
-4\lang (u\cdot G)^2-\frac{1}{4}G^2 \rang_N \biggl.\biggr\{\ 
(g_{\alpha\gamma}g_{\beta\delta}-g_{\alpha\delta}g_{\beta\gamma})
\nonumber\\
&-&2\;(g_{\alpha\gamma}u_\beta u_\delta-g_{\alpha\delta}u_\beta u_\gamma
-g_{\beta\gamma}u_\alpha u_\delta+g_{\beta\delta}u_\alpha u_\gamma)
\ \biggl.\biggr\}\ \Biggl.\Biggr],
\label{eqn:11}
\end{eqnarray}
where we define $(u\cdot G)^2 \equiv 
G_{\kappa\lambda}^a G_{\rho}^{a\;\lambda}\;u^\kappa u^\rho$.
By introducing the $u_\mu$, one can imagine 
uniformly moving uncorrelated nucleons ($p_\mu=M_N u_\mu$)
in nuclear matter, but in this case we set $u=(1,{\bf 0})$. 
The OPE expression 
for $T_{\mu\nu}$ is written as follows by 
the combination of Eq.(\ref{eqn:11}) and the Wilson coefficients
corresponding to each matrix element.  
\begin{eqnarray}
\lefteqn{\frac{1}{n!}\left(\frac{d}{dq^2}\right)^n
\frac{T^{{\rm OPE}}(q^2)}{q^2}\ 
\equiv\ \widehat{T}^{(n)\;{\rm OPE}}(q^2)}
\nonumber\\
&=&\frac{1}{3}\;\left[\ 
C_G^{(n)}(\xi)\left\{\ \lang\coup G^2\rang_N
-4\lang\coup {\cal ST}(G_{0\sigma}^a G_{0\sigma}^a)\rang_N\ \right\}\right.
\nonumber\\
&+&\left.\{D_1^{(n)}(\xi)-D_2^{(n)}(\xi)-D_3^{(n)}(\xi)\}
\ \lang \coup {\cal ST}(G_{0\sigma}^a G_{0\sigma}^a)\rang_N\ \right].
\label{eqn:12}
\end{eqnarray}
Here we define the dimensionless parameter 
as $\xi =  -q^2/4m_c^2$ ($m_c$; charmed quark mass) and 
$\rho=\xi/(1+\xi)$.
${\cal ST}$ means making the twist-2 operators
symmetric and traceless in its Lorentz indices. 
In Eq.(\ref{eqn:12}) each coefficient function is given 
using Gauss hypergeometric functions $_2F_1$ 
for arbitrary $q^2$ as follows:
\begin{eqnarray}
C_G^{(n)}(\xi)&=&-\frac{2^n(n+1)(n+3)!}{(2n+5)!!}(4m_c^2)^{-(n+2)}
(1+\xi)^{-(n+2)}\ _2F_1\left(n+2,-\frac{1}{2},n+\frac{7}{2};\rho\right)
\label{eqn:13}\\
D_1^{(n)}(\xi)&=&\frac{2^{n+3}(n+1)(n+1)!}{3(2n+3)!!}(4m_c^2)^{-(n+2)}
(1+\xi)^{-(n+2)} \nonumber\\
&\times& \left[\ 2\; _2F_1\left(n+2,\frac{1}{2},n+\frac{5}{2};\rho\right)
\right.
-\ \frac{2(n+2)}{1+\xi}\; 
_2F_1\left(n+3,\frac{1}{2},n+\frac{5}{2};\rho\right)
\nonumber\\
&&\qquad\qquad\ \ 
 \left.+\ \frac{3(n+2)^2}{(1+\xi)(2n+5)}\; 
_2F_1\left(n+3,\frac{1}{2},n+\frac{7}{2};\rho\right)\ \right]
\label{eqn:14}\\
D_2^{(n)}(\xi)&=&-\frac{2^{n+5}(n+1)(n+2)!}{3(2n+5)!!}(4m_c^2)^{-(n+2)}
(1+\xi)^{-(n+2)}
\nonumber\\
&&\times\left[\ _2F_1\left(n+2,\frac{1}{2},n+\frac{7}{2};\rho\right)\right.
-\frac{n+2}{2(1+\xi)}
\left.\; _2F_1\left(n+3,\frac{1}{2},n+\frac{7}{2};\rho\right)
\ \right]
\label{eqn:15}\\
D_3^{(n)}(\xi)&=&\frac{2^{n+3}(n+1)(n+1)!}{3(2n+5)!!}(4m_c^2)^{-(n+2)}
(1+\xi)^{-(n+2)} \nonumber\\
&\times& \left[\ (n+2)\; _2F_1\left(n+2,\frac{1}{2},n+\frac{7}{2};\rho\right)
\right.\left.+4(2n+5)\;_2F_1\left(n+2,\frac{1}{2},n+\frac{5}{2};\rho\right)
\ \right],\nonumber\\
&&\label{eqn:16}
\end{eqnarray} 
The Wilson coefficient of Eq.(\ref{eqn:13}) for scalar operator 
have already given in \cite{RRY2} and 
Eq.(\ref{eqn:14}), (\ref{eqn:15}) and (\ref{eqn:16})
are new contributions for twist-2 operator.
Eventually by equating Eq.(\ref{eqn:10}) 
and Eq.(\ref{eqn:12}) 
we obtain the moment sum rule expressed as the form 
of the $n$-th derivative 
with respect to $q^2$,
\begin{eqnarray}
\widehat{T}^{(n)\;{\rm ph}}(\xi\;;\; a,b)
= \widehat{T}^{(n)\;{\rm OPE}}(\xi).
\label{eqn:17}
\end{eqnarray}
The manipulation of the derivative ensures the enhancement 
of low energy part not to depend on the details 
of high energy part.
The vacuum sum rules have been utilized 
for investigation into the free state of charmonium
by Reinders $et\ al.$ \cite{RRY2} in moment sum rule 
and by Bertlmann \cite{RAB} in Borel sum rule. 
Furnstahl $et\ al.$ \cite{FHL} 
have studied the spectra of $J/\psi$
at finite temperature using both QCD sum rules. 
We summarize the well-known behavior of the moment sum rule for
the variations of $n$ and $q^2$.
\begin{itemize}
\item
The convergence of OPE side is worse with $n$ larger but
better with $q^2$ larger. 
\item
In contradiction to this behavior of OPE side,
the unwelcome contributions from the continuum in PH side  
grow with $q^2$ larger but decrease with $n$ larger.
\end{itemize}
We must choose the reliable 
stability region of moment sum rule for
the change of the both $n$ and $q^2$.
The moment sum rule for $T_{\mu\nu}$ is the method to investigate
the deviation from the properties in vacuum obtained from $\Pi^0$.
So we should adopt the same regions of  $n$ and $q^2$ as $\Pi_0$,
which can reproduce the nature of $J/\psi$
in the moment sum rule reasonably well.

\vskip0.5cm

\section{Moment sum rule for $\Pi_0$}
\setcounter{equation}{0}
\renewcommand{\theequation}{\arabic{section}.\arabic{equation}}
In this section we calculate the window of $n$ 
for various values of $\xi$ by applying the moment sum rule to
vacuum correlation function $\Pi_0$ \cite{RRY2}.

In OPE side $n$-th derivative for $\Pi^{(n)}_{0}$ 
is expressed as 
\begin{eqnarray}
\lefteqn{\frac{1}{n!}\left(\frac{d}{dq^2}\right)^n
\Pi^{\rm OPE}_{0}(q^2)\ 
\equiv\ \frac{\Pi^{(n)\;{\rm OPE}}_0(\xi)}{q^2}}
\nonumber\\
&=&\ \frac{1}{3}\;\left[\;
C_0^{(n)}(\xi)\ \{\;1+c_1^{(n)}(\xi)\;\alpha_s(\xi)\;\}+
C_G^{(n)}(\xi)\ \lang\coup G^2\rang_0\;\right],
\label{eqn:4.1}
\end{eqnarray}
where $C_I^{(n)}(\xi)$, $c_1^{(n)}(\xi)$ are given by
\begin{eqnarray}
C_0^{(n)}(\xi)&=&\frac{9}{4\pi^2}\frac{2^n(n+1)(n-1)!}{(2n+3)!!}
(4m_c^2)^{-n}(1+\xi)^{-n}
_2F_1\left(n,\frac{1}{2},n+\frac{5}{2};\rho\right),
\label{eqn:4.2}\\
c_1^{(n)}(\xi)&=&\frac{(2n+1)!!}{3\cdot 2^{n-1}n!}\frac{2n+3}{2(n+1)}
\frac{1}{_2F_1\left(n,\frac{1}{2},n+\frac{5}{2};\rho\right)}
\nonumber\\
&&\times \ \left[\;\pi-\left\{\frac{\pi}{3}
+\frac{1}{2}\left(\frac{\pi}{2}-\frac{3}{4\pi}\right)\right\}
\ \frac{1}{n+1}\ _2F_1\left(n,1,n+2;\rho\right) \right.
\nonumber\\
&&+\left. \frac{1}{3}\frac{1}{(n+1)(n+2)}
\left(\frac{\pi}{2}-\frac{3}{4\pi}\right)\ 
_2F_1\left(n,2,n+3;\rho\right)\right]
\nonumber\\
&&-\left(\frac{\pi}{2}-\frac{3}{4\pi}\right)
-2n\frac{{\rm ln}(2+\xi)}{\pi}\frac{2+\xi}{(1+\xi)^2}
\frac{_2F_1\left(n+1,\frac{1}{2},n+\frac{5}{2};\rho\right)}
{_2F_1\left(n,\frac{1}{2},n+\frac{5}{2};\rho\right)}.
\label{eqn:4.3}
\end{eqnarray}
In Eq.(\ref{eqn:4.1})
the OPE side of moment sum rule in vacuum $\Pi^{(n)}_0$
is the combination of the first term from a bare loop contribution 
and the second term replaced by the expectation values of the nucleon into
it of the vacuum.
On the other hand the relation between 
the $J/\psi$ mass of lowest resonance and $\Pi^{(n)}_{0}$
in PH side is given as
\begin{eqnarray}
\Pi^{(n)}_{0}(\xi)=\frac{9}{4}\frac{m_{J/\psi}^{2}}{g_{J/\psi}^{2}}
\frac{1}{(m_{J/\psi}^{2}
-\omega^{2})^{n+1}}[1+\delta_{n}],
\label{eqn:18}
\end{eqnarray}
where $m_{J/\psi}$ and $g_{J/\psi}$
are the parameters of the lowest-lying resonance and $\delta_{n}$
represents the contributions from higher resonances.
Moreover in $\Pi^{(n)}_0$ we add continuum contribution 
proportional to $1/4\pi\times (1+\alpha_s/\pi)$,
which we for simplicity assume to be 
constant value without the dependence of charmed quark mass.
From Eq.(\ref{eqn:18}) we eliminate the 
coupling parameter $g_{J/\psi}$ by taking ratios of the $n$-th moments
and the ($n-1$)-th moments.
Finally the bare mass of $J/\psi$ is derived from the following relation,
\begin{eqnarray}
m_{J/\psi} = \left[\omega^{2}+\frac{\Pi^{(n-1)}_{0}-\frac{1}{4\pi^{2}}
\left(1+\frac{\alpha_{s}}{\pi}\right)\frac{1}{n-1}\frac{1}
{(s_{0}-\omega^{2})^{n-1}}}{\Pi^{(n)}_{0}-\frac{1}{4\pi^{2}}
\left(1+\frac{\alpha_{s}}{\pi}\right)\frac{1}{n}
\frac{1}{(s_{0}-\omega^{2})^{n}}}\right]^{1/2}.
\label{eqn:19}
\end{eqnarray}

First we fix $\xi$ to be from $0.0$ to $3.0$ 
at $0.5$ intervals.
These values are equivalent 
to the magnitude of from $0$ to $4$ [GeV] in $\sqrt{-q^2}$.
In Fig.1 we show the $J/\psi$ bare mass determined from
Eq.(\ref{eqn:19}) for the change of $\xi$. Here we used
$s_0=3.6^2$ [GeV$^2$] adopted in \cite{RRY2}. 
We must read off the range of $n$ for each $\xi$ from the Figure 1.
The window of $n$ corresponds to find the region 
stabilizing the $J/\psi$ bare mass for the change of $n$.
Then we obtain the windows of $n$ for each $\xi$ as follows:
$n_{1}=2,3,4$ for $\xi=0.0$ ($-q^{2}=0.00$ [GeV$^2$]),
$n_{2}=3,4,5$ for $\xi=0.5$ ($-q^{2}=3.18$ [GeV$^2$]),
$n_{3}=4,5,6,7$ for $\xi=1.0$ ($-q^{2}=6.25$ [GeV$^2$]),
$n_{4}=5,6,7,8$ for $\xi=1.5$ ($-q^{2}=9.23$ [GeV$^2$]),
v$n_{5}=6,7,8,9$ for $\xi=2.0$ ($-q^{2}=12.1$ [GeV$^2$]),
$n_{4}=7,8,9,10$ for $\xi=2.5$ ($-q^{2}=14.9$ [GeV$^2$]),
$n_{5}=8,9,10,11$ for $\xi=3.0$ ($-q^{2}=17.6$ [GeV$^2$]).
These points seem to reproduce the bare mass reasonably well
for the experimental value $m_{J/\psi}=3.096$ [GeV].

\vskip0.5cm

\section{Numerical results}
\setcounter{equation}{0}
\renewcommand{\theequation}{\arabic{section}.\arabic{equation}}
By inserting the sets of $\xi$ and $n$ 
obtained in section 4 into Eq.(\ref{eqn:17}), we can determine
unknown parameters $a$ and $b$ simultaneously by fitting 
the left-hand side to the right-hand side.
Concretely the order of calculation is as follows;
At first we arbitrarily choose two points in the window of $n$ 
for the fixed $\xi$ and 
make a simultaneous equation for $a$ and $b$ by inserting two $n$
chosen.
We consider all such combinations
for each $\xi$ and take the average of $a$ solved for each combination.
Eventually the scattering length 
is easily obtained from $a$.

To calculate the scattering length we use the next values 
for other various parameters.
First of all, in PH side we adopt
$m_{J/\psi}=3.1\ [\mbox{GeV}]$,
$M_N=0.94\ [\mbox{GeV}]$ and 
$s_0=3.6^2\ [\mbox{GeV}^2]$ adopted in \cite{RRY2}.
The coupling is determined from the experimental value of 
$\Gamma_{J/\psi}^{e^+e^-}$ as follows:
\begin{eqnarray}
f_{J/\psi}^2=
\frac{3\Gamma_{J/\psi}^{ee}}{4\pi e_q^2 \alpha^2 m_{J/\psi}}
=1.7\times 10^{-2},
\label{eqn:20}
\end{eqnarray}
where $e_q$ is electric charge of quark ($e_c=2/3$ for charm quark) 
and $\alpha$ is the fine structure constant ($=1/137$).
Indeed this values of the coupling can be also determined 
by substituting the experimental
values of $J/\psi$ bare mass into Eq.(\ref{eqn:19}) oppositely. 
Then QSR for $\Pi^0$ reproduces well the experimental
values of the coupling \cite{RRY2}. 
For QCD Lagrangian parameters we use the following functions 
dependent on $\xi$ given in \cite{RRY2};
\begin{eqnarray}
\alpha_s(\xi)&=&\frac{\alpha_s(4m_c^2)}
{1+\frac{25}{12\pi}\;\alpha_s(4m_c^2)\;\mbox{ln}(1+\xi)}\ \ ,\qquad
\alpha_s(4m_c^2)\simeq 0.3
\label{eqn:21}\\
m_c(\xi)&=& 1.28 
\times \left[\ 1-\frac{\alpha_s(\xi)}{\pi}
\left\{\ \frac{2+\xi}{1+\xi}\;\mbox{ln}(2+\xi)-2\;\mbox{ln}2
\right\}\ \right]\ [\;\mbox{GeV}\;]
\label{eqn:22}
\end{eqnarray}
In OPE side we determine the nucleon matrix elements as follows:
\begin{eqnarray}
\lang\coup G^2\rang_N &=& -(1.222\pm 0.282)\ \ [\;\mbox{GeV}^2\;]
\label{eqn:23}\\
\lang\coup {\cal ST}(G^a_{0\sigma}G^a_{0\sigma})\rang_N
&=& -(0.094\pm 0.010)\ \ [\;\mbox{GeV}^2\;]
\label{eqn:24}
\end{eqnarray}
The scalar part is evaluated from the $\pi$-$N$ sigma term \cite{JCFG,HK}. 
The twist-2 part is determined from the gluon
distribution function of a nucleon, which is obtained by leading
order parametrization to the experimental data 
of deep inelastic scattering \cite{JCFG,XJi}. 

We list the results in the case of scalar operator only
and the case involving twist-2 operator in Table 1.

\begin{center}
\begin{table}
\begin{center}
\begin{tabular}{|c|c|c|c|c|c|c|c|}\hline
$\xi$ & $0.0$ &
        $0.5$ &
        $1.0$ &
        $1.5$ &
        $2.0$ &
        $2.5$ &
        $3.0$ \\ \hline
$-\;a_V^{\rm scalar}$ [fm] &
  0.091 &
  0.068 &
  0.070 &
  0.063 &
  0.059 &
  0.057 &
  0.055 \\ \hline
$\delta m_V^{\rm scalar}$ [MeV] &
    5.3 &
    3.9 &
    4.0 &
    3.6 &
    3.4 &
    3.3 &
    3.2  \\ \hline\hline
$-\;a_V^{\rm twist-2}$ [fm] &
  0.120 &
  0.090 &
  0.092 &
  0.083 &
  0.078 &
  0.075 &
  0.073 \\ \hline
$\delta m_V^{\rm twist-2}$ [MeV] &
    6.9 &
    5.2 &
    5.3 &
    4.8 &
    4.5 &
    4.3 &
    4.2  \\ \hline
\end{tabular}
\end{center}
\caption{$J/\psi$-$N$ scattering length and the mass shift of $J/\psi$
in the case of the scalar
operator only and the scalar $+$ twist-2 operator 
at normal matter density $\rho_N=0.17\ [\mbox{fm}^{-3}]$. } 
%\protect{\cite{}}
\label{table:table1}
\end{table}
\end{center}

\vskip0.5cm

\section{Concluding remarks}
\setcounter{equation}{0}
\renewcommand{\theequation}{\arabic{section}.\arabic{equation}}
The direct application of moment sum rule to 
the forward $J/\psi$-$N$ scattering amplitude supplies us the fascinating 
result for the $J/\psi$-$N$ interaction. 
That is, the $J/\psi$-$N$ scattering length $a_{J/\psi}$ indicates
negative value (about $- 0.1$ fm). This result suggests that
the attractive $J/\psi$-$N$ interaction is not sufficient to
form a bound state with one nucleon, but 
it could make a bound state with nuclei.
The absolute value is certainly smaller 
than the typical hadronic size $1$ fm 
and the scattering length of light vector meson-
$N$ systems ($a_{\rho} \simeq - 0.47$, $a_{\omega} \simeq - 0.41$, 
$a_{\phi} \simeq - 0.15$) \cite{KH}, 
but the experimental creation of $J/\psi$ at the threshold 
would lead to formation of a bound state inside a heavy nucleus.  
Our result is smaller than those obtained recently by 
S.J.Brodsky et al. \cite{BM} and 
G.F.de T\'eramond et al. \cite{TER} in the 
QCD sum rule approach \cite{MEP}.
Their method is based on the on-shell calculation 
on charmed quark mass ($q^2=0$).

In this study we newly calculated the Wilson coefficients for 
twist-2 gluon operators (dimension-4)
in the form with quark mass explicitly.
The nucleon matrix elements of twist-2 gluon operators is 
about 1/10 times as large as that of the scalar part, but 
the total contribution with the Wilson coefficient
makes the absolute value of scalar part larger about $30\%$.

From $a_{J/\psi}$, we can estimate a total cross section 
($\sigma_{J/\psi}$ $=$ $4\pi a_{J/\psi}^2$).
The result is about $1.26$ mb at the threshold. The contribution 
from the elastic channel
corresponds to $20\,\%$ 
of the nuclear absorption cross section derived 
from the experimental data,
$\sigma_{nab}$ $=$ $7.3$ mb.

Next in the linear density 
approximation we can calculate $J/\psi$ mass shift from $a_{J/\psi}$. 
The result gives very small decrease of mass
(about $-4$ to $-7$ MeV), 
about $0.1$ to $0.2\%$ at normal matter density.
Since the slight mass shift is of the order of MeV, 
the change is sufficiently
larger than the leptonic decay width of the order keV. 
So we consider $J/\psi$ is a good probe for the observation of medium effect.  

Recently just before I submit this paper,
Klingl {\it et al.} reported the mass shift of about $-20$ MeV
($0.6\%$) for $J/\psi$ in nuclear matter \cite{KKLMW}. 
The pattern of derivation is similar to 
the case for light vector meson \cite{HL,KH}.

\vspace*{1in}

\centerline{\large{\bf Acknowledgements}}
I would like to thank Y. Koike for helpful and fruitful discussions and 
T. Hatsuda and T. Matsui for useful comments and information.
I also thank K. Itakura for a careful and critical reading of the manuscript 
and useful discussions.

	\newpage

\begin{center}
{\bf Figure Captions}
\end{center}
\vskip .5cm

\noindent {\bf Fig.~1} Stability region in n for various
values of $\xi$. For comparison the experimental mass values
have also been indicated.\\

\noindent {\bf Fig.~2} $J/\psi$-$N$ scattering length 
determined for various values of $\xi$
and the stability region of n obtained from Fig.1.

\newpage

\begin{figure}[h]
\begin{center}
\epsfile{file=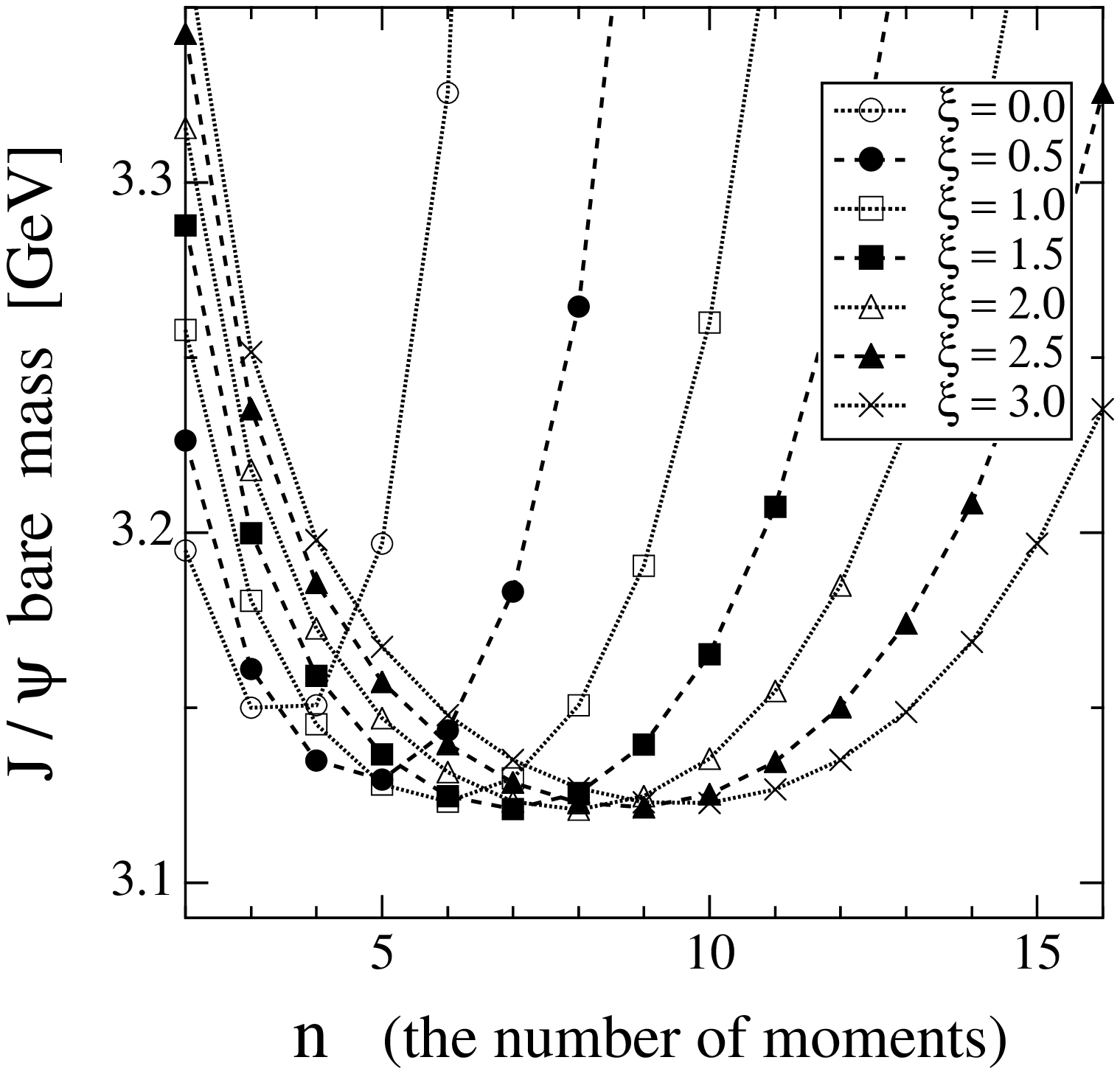,scale=0.8}
\end{center}
%\protect{\cite{}}
\label{figure:fig1}
\end{figure}

\newpage

\begin{figure}[h]
\begin{center}
\epsfile{file=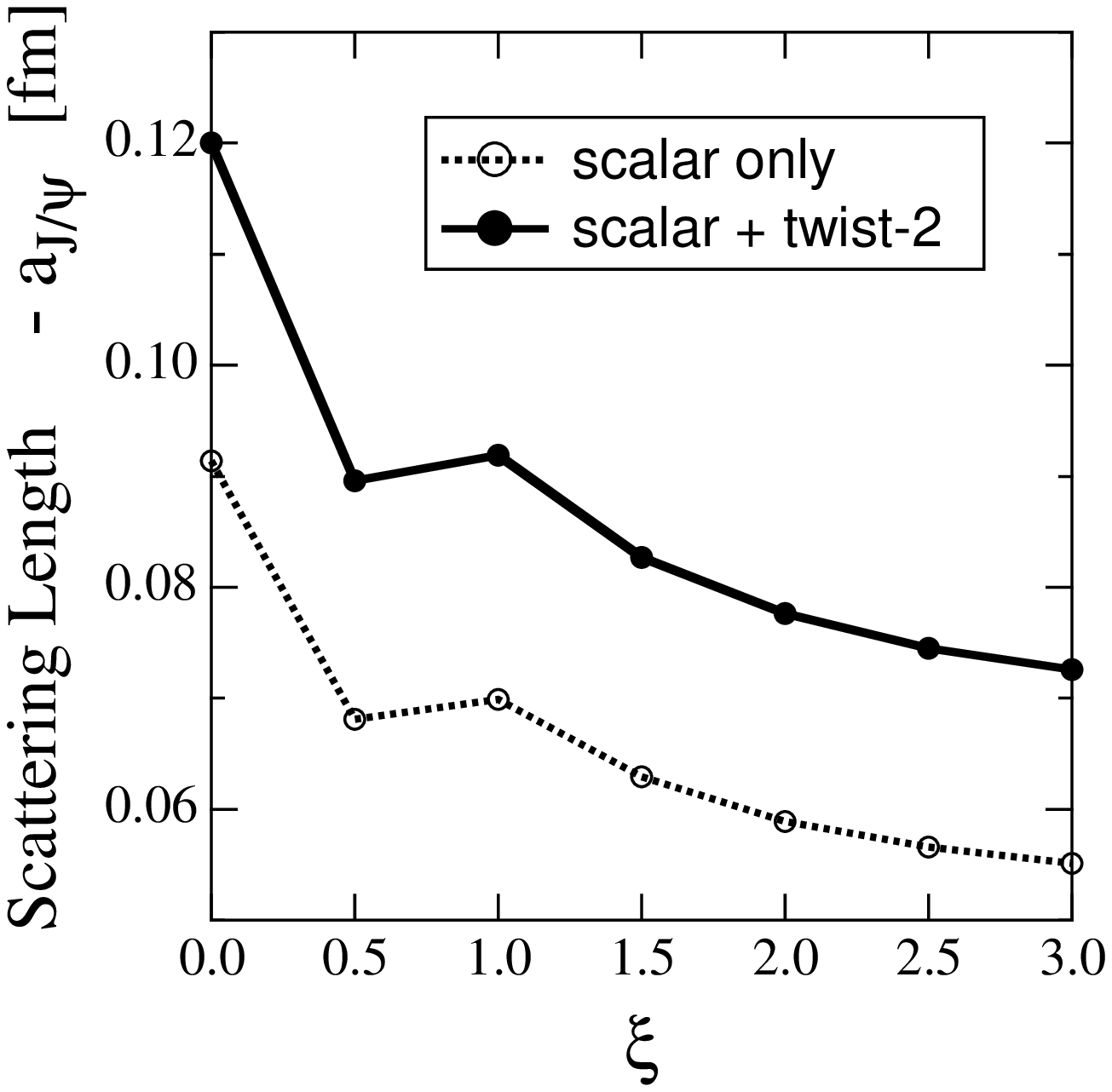,scale=0.87}
\end{center}
%\protect{\cite{}}
\label{figure:fig2}
\end{figure}

\end{document}